\documentclass[12pt,epsf,amssymb,ulem,qsymbols]{article}
\usepackage{tabularx}
\usepackage{array}
\usepackage{graphics}
\usepackage{graphicx}
\usepackage{psfrag}
\usepackage{epsfig}
\usepackage{amsmath}
\usepackage{amssymb}
\usepackage{ulem}
\usepackage{setspace}
\usepackage{rotating}
\usepackage{colortbl}
\usepackage{tabularx}
\usepackage{longtable}
\usepackage{multirow}
\makeatletter

\usepackage{verbatim}

\setlongtables

\newcommand{\msbar}{{\overline{\rm MS}}}

\newcommand{\bea}{\begin{eqnarray}}
\newcommand{\eea}{\end{eqnarray}}
\newcommand{\beq}{\begin{equation}}
\newcommand{\eeq}{\end{equation}}
\newcommand{\ec}{\end{center}}
\newcommand{\bc}{\begin{center}}
\newcommand{\gev}{{\rm GeV}}
\newcommand{\mev}{{\rm MeV}}

\newcommand{\pdir}{p\kern -5.2pt\raise 0.2ex\hbox {/}}

\newcommand{\vdir}{v\kern -5.75pt\raise 0.15ex\hbox {/}}
\newcommand{\kdir}{k\kern -5.75pt\raise 0.15ex\hbox {/}}
\newcommand{\epsdir}{\epsilon\kern -5.0pt\raise 0.15ex\hbox {/}}
\newcommand{\bvdir}{\bar{v}\kern -5.75pt\raise 0.15ex\hbox {/}}
\newcommand{\Ddir}{D\kern -7.75pt\raise 0.20ex\hbox {/}}
\newcommand{\Adir}{A\kern -7.75pt\raise 0.20ex\hbox {/}}
\newcommand{\ldir}{l\kern -5.0pt\raise 0.2ex\hbox{/}}
\newcommand{\varepsdir}{\varepsilon\kern -5.5pt\raise 0.15ex\hbox{/}}

\newcommand{\nn}{\nonumber}
\makeatother

\begin{document}
\thispagestyle{empty} 
\begin{flushright}
\begin{tabular}{l}
LPT 11-119\\
RM3-TH/12-1
\end{tabular}
\end{flushright}
\begin{center}
\vskip 0.8cm\par
{\par\centering \textbf{\LARGE  
\Large \bf $D$-meson decay constants and a check of }}\\
\vskip .35cm\par
{\par\centering \textbf{\LARGE  
\Large \bf factorization in non-leptonic $B$-decays }}\\
\vskip 1.05cm\par
{\scalebox{.8}{\par\centering \large  
\sc Damir Be\v{c}irevi\'c$^a$, Vittorio Lubicz$^{b,c}$, Francesco Sanfilippo$^{a,d}$}
{\par\centering \vskip 0.22 cm\par}
{\scalebox{.8}{\par\centering \large  
\sc Silvano Simula$^{c}$ and Cecilia Tarantino$^{b,c}$ }}
{\par\centering \vskip 0.65 cm\par}
{\sl 
$^a$~Laboratoire de Physique Th\'eorique (B\^at.~210)~\footnote{Laboratoire de Physique Th\'eorique est une unit\'e mixte de recherche du CNRS, UMR 8627.}\\
Universit\'e Paris Sud, F-91405 Orsay-Cedex, France.}\\
{\par\centering \vskip 0.25 cm\par}
{\sl 
$^b$~Dipartimento  di Fisica, Universit\`a RomaTre, \\
Via della Vasca Navale 84, I-00146 Roma, Italy.} \\
{\par\centering \vskip 0.25 cm\par}
{\sl 
$^c$~INFN, Sezione di RomaTre, \\
Via della Vasca Navale 84, I-00146 Roma, Italy.}\\ 
{\par\centering \vskip 0.25 cm\par}
{\sl 
$^d$~INFN, Sezione di Roma, \\
Piazzale Aldo Moro 5, I-00185 Roma, Italy.}\\ 
{\vskip 1.05cm\par}}
\end{center}

\vskip 0.55cm
\begin{abstract}
We compute the vector meson decay constants $f_{D_{(s)}^\ast}$ from the simulation of twisted mass QCD on the lattice with $N_{\rm f}=2$ dynamical quarks. When combining these values with the pseudoscalar $D_{(s)}$-meson decay constants, we were able (i) to show that the heavy quark spin symmetry breaking effects with the charm quark are large, $f_{D^\ast_s}/f_{D_s}=1.26(3)$, and (ii) to check the factorization approximation in a few specific $B$-meson non-leptonic decay modes. Besides our main results, ${f_{D^\ast}}  = 278\pm 13\pm 10$~MeV, and ${f_{D^\ast_s}}  = 311\pm 9$~MeV, other phenomenologically interesting results of this paper are: 
 ${f_{D^\ast_s}/f_{D^\ast}}  = 1.16\pm 0.02\pm 0.06$, ${f_{D^\ast_s}/f_{D}}  = 1.46\pm 0.05\pm 0.06$, and ${f_{D_s}/f_{D^\ast}}  = 0.89\pm 0.02\pm 0.03$. Finally, we correct the value for $B(B^0\to D^+\pi^-)$ quoted by PDG, and find $B(B^0\to D^+\pi^-) =  \left( 7.8 \pm 1.4 \right) \times 10^{-7}$. Alternatively, by using the ratios discussed in this paper, we obtain $B(B^0\to D^+\pi^-) =  \left( 8.3 \pm 1.0 \pm 0.8 \right) \times 10^{-7}$.

\end{abstract}
%\vskip 2.2 cm 
\newpage
\setcounter{page}{1}
\setcounter{footnote}{0}
\setcounter{equation}{0}
%%%%%%%%%%%%%%%%%%%%%%%%%%%%%%%%%%%%%%%%
\noindent

\renewcommand{\thefootnote}{\arabic{footnote}}

\setcounter{footnote}{0}
%%%%%%%%%%%  Section 1
\section{\label{sec-0}Introduction}
Vector meson decay constants are important ingredients in the particle physics phenomenological description of various processes~\cite{pheno}. They are particularly handy when checking on the validity of the  factorization approximation in non-leptonic decay channels involving a vector meson. Of particular interest are the $ B^0$-meson non-leptonic modes in which one of the two charged mesons in the final state is $D^{(\ast)}_{(s)}$ and the other one is a light meson. Factorization is expected to work quite well for this class of decay modes~\cite{factorization1}. In the specific cases such as  $B^0 \to D_{(s)}^{(\ast)-}\pi^+$, the factorization was shown to be exact in the limit of infinitely heavy quark mass~\cite{factorization2}. Away from that limit and in other similar situations, such as $B^0 \to D_{(s)}^{(\ast)+}\pi^-$, the factorization is an assumption~\cite{ff2}. While it is difficult to check on the extent to which the factorization approximation works for the absolute values of the branching fractions, it is relatively simple to do it when considering the ratios of various modes. $B$-factory experiments at BaBar and Belle provided us with many accurate measurements of the non-leptonic $B$-decays, of which particularly interesting are the measured $B^0 \to D_{(s)}^{(\ast)+}\pi^-$, and $B^0 \to D_{(s)}^{(\ast)+}D^-$ modes~\cite{PDG}.Ä
For example, in the naive factorization the amplitude for the $B^0\to D^+\pi^-$ decay writes,
\bea
A_{\rm fact}= - \frac{G_F}{ \sqrt{2}} V_{ub} V_{cd}^\ast \left[c_2(m_b)+\frac{1}{ N_c}c_1(m_b)\right] \langle D^+ \vert \bar c \gamma_\mu^L d\vert 0\rangle \langle \pi^- \vert \bar b \gamma^\mu_L u\vert B^0\rangle \,,
\eea
where the Wilson coefficients $c_{1,2}(\mu)$ are known to next-to-leading order (NLO) in QCD perturbation theory.~\footnote{More specifically, from ref.~\cite{buras}, one reads: $c_1^{\rm NLO}(m_b)=-0.285(14)$ and $c_2^{\rm NLO}(m_b)=1.132(8)$.} The hadronic matrix elements, instead, are non-perturbative quantities and need to be accurately evaluated on the lattice. The amplitudes for the other modes can be written similarly, and it is easy to see that in the suitable ratios the semileptonic hadronic matrix element $\vert  \langle \pi^- \vert \bar b \gamma^\mu_L u\vert B^0\rangle\vert$ cancels out, leaving only the ratios of {\sl vacuum}-to-$D^{(\ast)}_{(s)}$ meson matrix elements to be computed~\cite{ff2}. This is a favorable situation for the lattice QCD studies that will be explored in this paper. The factorization approximation is shown to work for $B^0 \to D_{(s)}^{(\ast)-}\pi^+$. 
We will present the results of our computation of the $D^\ast$ and $D_s^\ast$ meson decay constants ($f_{D^\ast}$ and $f_{D_s^\ast}$), that can be combined with the pseudoscalar ones ($f_D$ and $f_{D_s}$) and check the extent to which the factorization approximation provides the adequate description of the ratios of several non-leptonic $B$ decay modes. From the lattice QCD point of view, particularly appealing is the computation of $f_{D_s^\ast}/f_{D_s}$ because both the physical charm and strange quarks are directly accessible in the simulations, and the final physical result requires only a smooth extrapolation in the sea light quark mass and to the continuum limit. 

Our computation of $f_{D^\ast}$ and $f_{D_s^\ast}$ is made on the ensembles of gauge field configurations produced by  the  European Twisted Mass Collaboration (ETMC) at four different lattice spacings and for several light sea quark masses~\cite{Boucaud:2008xu}, by using the twisted mass QCD on the lattice (LtmQCD)~\cite{fr}. In this way the chiral and continuum extrapolations are well under control. The results presented here are unquenched, with $N_{\rm f}=2$ flavors of dynamical mass-degenerate light quarks. We briefly discuss the corresponding pseudoscalar meson decay constants as well that will help us answering to the  question about the size of the heavy quark spin symmetry breaking effect in the heavy-light mesons when the heavy quark is charm. In our work this means that we want to measure how far from the static limit we actually are. In the static limit, $\displaystyle{\lim_{m_c\to \infty}}(f_{D^\ast}/f_D)=1$. As we shall see, our results suggest that this ratio is considerably larger than one. 

The remainder of this paper is organized as follows. In Sec.~\ref{sec-2} we remind the reader of the definition of the pseudoscalar and vector meson decay constants, and discuss the peculiarity of their lattice determination in the framework of LtmQCD. In Sec.~\ref{sec-3} we summarize the details of the lattices used in our analysis, present our results as extracted from the correlation functions studied on each of the available data-sets, and after the chiral and continuum extrapolation we present our physical results. Using our results, we then compare the ratios of several non-leptonic decay modes measured experimentally with the results of the factorization approximation. In Sec.~\ref{sec-5} we briefly conclude.

\section{\label{sec-2}Open charm mesons and their decay constants}

The decay constants of charmed pseudo-scalar and vector mesons are defined through the following matrix elements,
\bea \label{def1}
&&\langle 0\vert \bar c (0) \gamma_\mu \gamma_5 q(0)
\vert D_q (p) \rangle = f_{D_q} p_\mu \;, \nonumber \\
&& \\
&&\langle 0\vert \bar c (0) \gamma_\mu q(0)
\vert D_q^\ast (p,\lambda) \rangle = f_{D^\ast_q} m_{D^\ast_q} e_\mu^\lambda \,, \nonumber
\eea
where the index $q$ stands for either the strange or the light $u/d$-quark. We assume isospin symmetry and do not distinguish between the $u$- and $d$-quark mass.  In the above expressions, $p$ and $e_\mu^\lambda$ are the meson momentum and the vector meson polarization, respectively. 

Instead of the axial current matrix element, in LtmQCD, it is far more convenient to extract the pseudoscalar meson decay constant from the matrix element of the pseudoscalar density. At the maximal twist, the renormalization constant of the quark mass and the pseudoscalar densities cancel exactly, and therefore 
\bea
(\mu_q+\mu_c) \langle 0\vert \bar c (0) \gamma_5 q(0)
\vert D_q (p) \rangle = f_{D_q}   m_{D_q}^2\,,
\eea
with $\mu_q$ and $\mu_c$ being the light and charm bare quark masses respectively. The results of the computations of $f_{D_q}$ in LtmQCD at the maximal twist were first reported in ref.~\cite{Blossier:2009bx}. They were later corroborated  in ref.~\cite{Dimopoulos:2011gx}, i.e. after including the results of simulations made at finer lattice spacing, corresponding to $\beta=4.2$. In this paper we will focus onto the vector meson decay constants, $f_{D_q^\ast}$, the results of which we present together with the pseudoscalar ones, as to better emphasize the fact that the charm is indeed far from the static heavy quark limit. In contrast to the pseudoscalar case, the vector meson decay constant cannot be extracted  without  explicitly accounting for a renormalization factor~\cite{fr}. The relevant constant for the physical vector current in LtmQCD is $Z_A(g_0^2)$, and its value for all our lattices has been computed non-perturbatively in ref.~\cite{Constantinou:2010gr}.

The access to the vector meson mass ($m_{D^\ast_q}$) and its decay constant ($f_{D^\ast_q}$) on the lattice is made through the study of the large time separation in the two-point correlation function involving the vector current, $V_i= Z_A(g_0^2) \bar c \gamma_i q$, namely
\bea
\label{r1}
&& C_{VV}(t)  =  \langle {\displaystyle \sum_{\vec x} }   V_{i}(\vec x; t) V^\dagger_{i}(0; 0) \rangle = - {\rm Tr}\left[ S_c(0,0;\vec x,t)\gamma_i S_q(\vec x,t;\vec 0,0) \gamma_i\right] \nn\\
&&\qquad \xrightarrow[]{\displaystyle{ t\gg 0}} \; \frac{\cosh[  m_{D^\ast_q} (T/2-t)]}{ m_{D^\ast_q} }  \left| \langle 0\vert V_i(0)
\vert D_q^\ast (\vec 0, \lambda) \rangle \right|^2 e^{- m_{D^\ast_q} T/2}
\;.
\eea
In the above expression $S_{q(c)}(x;0)$ stands for the light (charm) quark propagator. Recently, the authors of ref.~\cite{Endress:2011jc} noted that the use of stochastic source propagators in the computation of vector meson properties is not as advantageous as it is with the pseudoscalar mesons. That was in fact already observed in refs.~\cite{Dimopoulos:2011cf,Jansen:2009hr}, but in ref.~\cite{Endress:2011jc} the authors show that the extraction of the light vector meson properties from the lattice can be improved if several stochastic sources in the propagator inversion are used. 
\section{\label{sec-3}Lattice details and results}

This section contains the main details of our computation, including the discussion of the extrapolation procedure that leads to the phenomenologically relevant results. 
\begin{table}[h!!]
\centering 
{\scalebox{.93}{\begin{tabular}{|c|cccccc|}  \hline \hline
{\phantom{\huge{l}}}\raisebox{-.2cm}{\phantom{\Huge{j}}}
$ \beta$& 3.8 &  3.9  &  3.9 & 4.05 & 4.2  & 4.2    \\ 
{\phantom{\huge{l}}}\raisebox{-.2cm}{\phantom{\Huge{j}}}
$ L^3 \times T $&  $24^3 \times 48$ & $24^3 \times 48$  & $32^3 \times 64$ & $32^3 \times 64$& $32^3 \times 64$  & $48^3 \times 96$  \\ 
{\phantom{\huge{l}}}\raisebox{-.2cm}{\phantom{\Huge{j}}}
$ \#\ {\rm meas.}$& 240 &  240$^\ast$ & 240 & 240 & 240 & 96  \\ \hline 
{\phantom{\huge{l}}}\raisebox{-.2cm}{\phantom{\Huge{j}}}
$\mu_{\rm sea 1}$& 0.0080 & 0.0040 & 0.0030 & 0.0030 & 0.0065 &  0.0020   \\ 
{\phantom{\huge{l}}}\raisebox{-.2cm}{\phantom{\Huge{j}}}
$\mu_{\rm sea 2}$& 0.0110 & 0.0064 & 0.0040 & 0.0060 &   &     \\ 
{\phantom{\huge{l}}}\raisebox{-.2cm}{\phantom{\Huge{j}}}
$\mu_{\rm sea 3}$&  & 0.0085 &  & 0.0080 &   &     \\ 
{\phantom{\huge{l}}}\raisebox{-.2cm}{\phantom{\Huge{j}}}
$\mu_{\rm sea 4}$&  & 0.0100 &  &   &   &     \\   \hline 
{\phantom{\huge{l}}}\raisebox{-.2cm}{\phantom{\Huge{j}}}
$a \ {\rm [fm]}$&   0.098(3) & 0.085(3) & 0.085(3) & 0.067(2) & 0.054(1) & 0.054(1)      \\ 
{\phantom{\huge{l}}}\raisebox{-.2cm}{\phantom{\Huge{j}}}
$Z_A (g_0^2)$~\cite{Constantinou:2010gr,Blossier:2010cr}& 0.746(11) & 0.746(6) & 0.746(6)  & 0.772(6) & 0.780(6)& 0.780(6) \\ 
{\phantom{\huge{l}}}\raisebox{-.2cm}{\phantom{\Huge{j}}}
$\mu_{s}$& 0.0194(12)  &0.0177(11)  &0.0177(11)  & 0.0154(10) & 0.0129(10) & 0.0129(10)  \\ 
{\phantom{\huge{l}}}\raisebox{-.2cm}{\phantom{\Huge{j}}}
$\mu_{c}$& 0.2331(82)  &0.2150(75)  &0.2150(75)   & 0.1849(65) & 0.1566(55) & 0.1566(55)  \\ 
 \hline \hline
\end{tabular}}}
%%%%%%%%%%%%%%
{\caption{\footnotesize  \label{tab:1} Summary of the details about the lattice ensembles used in this work. The asterisk in the number of gauge field configurations used from the simulations at $\beta =3.9$ is there to indicate that the data-set with $\mu_{\rm sea 1}=0.0040$ contains $480$ configurations, while the sets with other values of $\mu_{\rm sea}$ contain $240$. Data obtained at different $\beta$'s are rescaled by using the Sommer parameter $r_0/a$, and the overall lattice spacing is fixed by matching $f_\pi$ obtained on the lattice with its physical value, leading to  $r_0= 0.440(12)$~fm (c.f. ref.~\cite{Blossier:2010cr}). Strange and charm bare quark masses, $\mu_s$ and $\mu_c$ respectively, are obtained as discussed in ref.~\cite{Blossier:2010cr}. All quark masses are given in lattice units.}}
\end{table}
\subsection{Lattices and correlation functions}  
Details about the sets of gauge field configurations with $N_{\rm f}=2$ dynamical quarks that are used in this work are summarized in table~\ref{tab:1}. 

\begin{figure}[h!!]
\begin{center}
\begin{tabular}{@{\hspace{-0.25cm}}c}
\epsfxsize12.2cm\epsffile{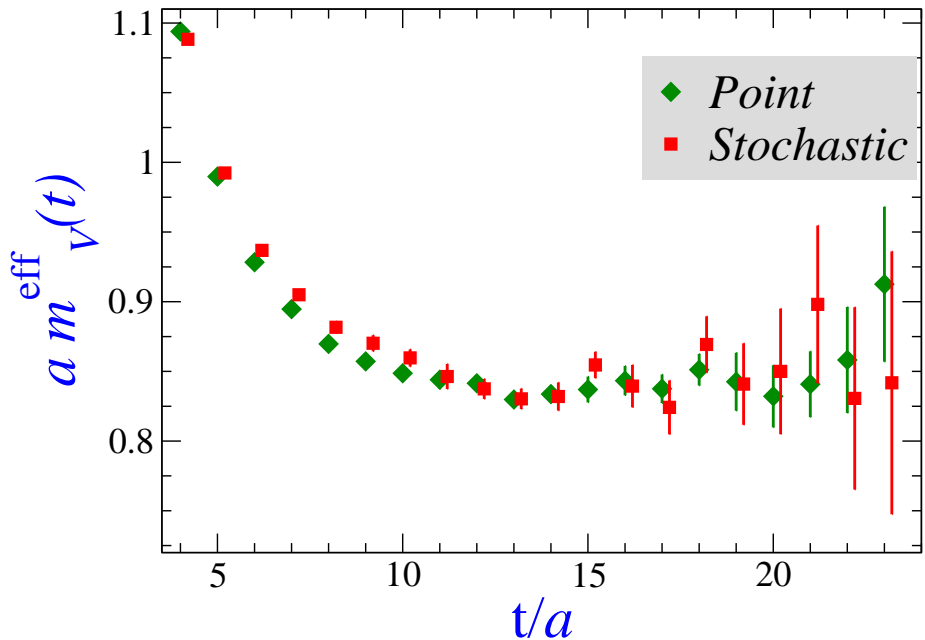}    \\
\end{tabular}
\vspace*{-.1cm}
%%%%%%%%%%%%%%%%%%%%%%%%%%%%%%%%%%%%%%%%%%%%%%%%%%%%%%%%%%%%%%%%%%
\caption{\label{fig:1}{\footnotesize 
Comparison of the effective masses of the heavy-light vector mesons obtained from the correlation function (\ref{r1}) computed by combining either point source propagators or stochastic source propagators. Illustration  is provided for the data sets obtained at $\beta=3.9$ and with $\mu_{\rm sea}=0.0064$. 
 } }
%%%%%%%%%%%%%%%%%%%%%%%%%%%%%%%%%%%%%%%%%%%%%%%%%%%%%%%%%%%%%%%%%%
\end{center}
\end{figure}

Like in the case of the light-light vector currents, in the corresponding heavy-light correlation functions we observe that the point source propagators  lead to the equally good results for the decay constants and hadron masses as those obtained by using the stochastic source propagators. We did not attempt using more stochastic sources. We merely note that the quality of the signal for the two-point correlation functions obtained by two kinds of propagators is the same and the effective mass plateaus remain indistinguishable, c.f. fig.~\ref{fig:1}.~\footnote{The behavior of $m_V^{\rm eff}(t)$ obtained with the point source propagators is less stable even if the error bars are smaller, while  the time dependence of the effective mass is more stable from the correlation functions with stochastic propagators but with slightly larger errors. As a result the  vector meson masses from the correlators with either kind of propagators are indistinguishable. } The effective mass is the solution of,
\bea
{\cosh\left[ m_V^{\rm eff}(t) \left( {\displaystyle{T\over 2}} - t\right)\right] \over \cosh\left[ m_V^{\rm eff}(t) \left( {\displaystyle{T\over 2}} - t -1\right)\right] }  = {C_{VV}(t)\over C_{VV}(t+1)}\,.
\eea  
In the following the results we present are obtained by using the stochastic propagators~\cite{Boucaud:2008xu}. 
%%%%%%%%%%%%%%%%%%%%%%%%%%%%%%%%%%%%%%%%%%%%%%%%%%%%%%%%%%%

\begin{figure}[h]
\begin{center}
\begin{tabular}{@{\hspace{-0.25cm}}c}
\epsfxsize12.2cm\epsffile{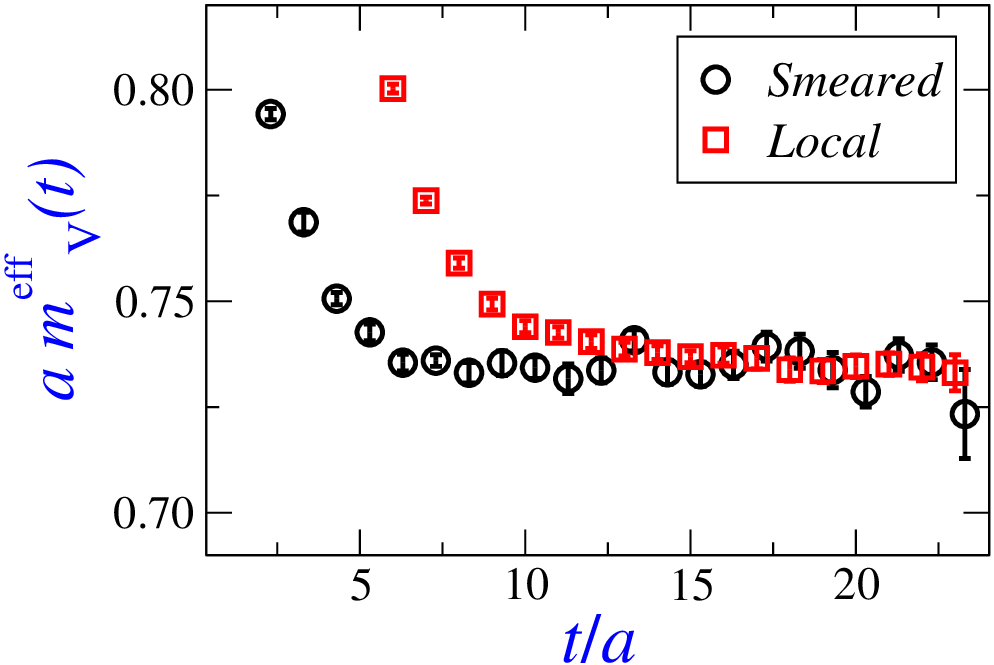}    \\
\end{tabular}
\vspace*{-.1cm}
%%%%%%%%%%%%%%%%%%%%%%%%%%%%%%%%%%%%%%%%%%%%%%%%%%%%%%%%%%%%%%%%%%
\caption{\label{fig:2}{\footnotesize 
Effective mass of the heavy-light vector meson, extracted from the two-point correlation function~(\ref{r1}) obtained with either the local vector currents or by implementing Gaussian smearing. At larger time separations we see that the two results are totally compatible. Illustration is provided on the data at $\beta=3.9$ and $\mu_{\rm sea}=0.0100$. 
 } }
%%%%%%%%%%%%%%%%%%%%%%%%%%%%%%%%%%%%%%%%%%%%%%%%%%%%%%%%%%%%%%%%%%
\end{center}
\end{figure}

Another worry when dealing with heavy quarks on the lattice is to make sure that the lowest lying state has been well isolated at moderately large time separations between the interpolating operators. In refs.~\cite{Blossier:2009bx,Dimopoulos:2011gx} the values for $m_{D_q}$ and $f_{D_q}$ were obtained from the correlation functions with the local sources. We checked that by using Gaussian smearing (see e.g. ref.~\cite{Gattringer:2010zz}) the plateaus indeed become larger but the error bars on the resulting decay constants remain the same. Furthermore the central values for the pseudoscalar masses and decay constants remain unchanged (within the statistical accuracy). A similar situation holds true in the case of vector mesons, i.e. the effective mass indeed exhibits plateau `earlier' when the smeared source operators are used. At moderately large time separations, where the extraction of the masses and decay constants is made, the effective masses obtained from correlation functions with local and with smeared interpolating operators coincide. This point is illustrated in fig.~\ref{fig:2}. In what follows we will quote as our main results those obtained by using local operators with stochastic sources. We checked that our final results remain unaltered when the smeared correlation functions are combined with the local ones.
%%%%%%%%%%%%%%%%%%%%%%%%%%%%%%%%%%%%%%%%%%%%%%%%%%%%%%%%%%%
\subsection{ $f_{D_s^\ast}$ and  $f_{D_s^\ast}/ f_{D_s}$}
With the above comments in mind, we now present our results for the charm-strange pseudoscalar and vector mesons. Since the ensembles of gauge field configurations used in this work are obtained with $N_{\rm f}=2$ light quarks, the results presented in this subsection are only partially unquenched. The strange valence or charmed quarks are directly accessible from our lattices but they do not have their ``sea"-quark counterparts. Only the light sea quarks (denoted in what follows as $m_q$) have been included in the QCD vacuum fluctuations. The argument that the strange quark mass is too heavy to significantly contribute to the QCD vacuum fluctuations seems to be confirmed by the fact that the $f_{D_s}$ decay constants obtained from the simulations with $N_{\rm f}=2$~\cite{Dimopoulos:2011gx,fDnf=2} and $N_{\rm f}=2+1$~\cite{fDnf=3} are essentially equal. Of course such a comparison is obscured by the fact that different lattice QCD actions have been used in obtaining the respective results. As far as the charm quark is concerned, none of the currently available lattice results on charm physics include the charm quark in the sea. That is likely to change in the near future as the ETMC and MILC are continuously producing the gauge field configurations with $N_{\rm f}=2+1+1$ dynamical quark flavors, i.e. with both the strange and charm quarks in the sea~\cite{Nf=2+2,Bazavov:2010ru}. 
\begin{table}[t!!]
\centering 
{\scalebox{1.}{\begin{tabular}{|c c c|c|c|c|c|}  \hline
{\phantom{\huge{l}}} \raisebox{-.2cm} {\phantom{\huge{j}}}
$\beta$ &  \qquad$L$ & \qquad $\mu_{\rm sea }$   & $ m_{D_s}$            & $ m_{D^\ast_s} $       & $f_{D_s}$          & $f_{D^\ast_s}$       \\ \hline\hline
{\phantom{\huge{l}}} \raisebox{-.2cm} {\phantom{\huge{j}}}
3.80 &  \qquad 24 & \qquad$\mu_{\rm sea 1}$ & 1.773(54) &  2.008(59) & 0.280(8)   & 0.352(15)\\ 
{\phantom{\huge{l}}} \raisebox{-.2cm} {\phantom{\huge{j}}}
   &&\qquad$\mu_{\rm sea 2}$    & 1.771(54) & 2.005(58) & 0.279(9)  & 0.351(15)\\ \hline
{\phantom{\huge{l}}} \raisebox{-.2cm} {\phantom{\huge{j}}}
3.90 & \qquad 24 &\qquad$\mu_{\rm sea 1}$ & 1.806(44) &  2.023(46) &  0.265(7)  &  0.326(10)\\
{\phantom{\huge{l}}} \raisebox{-.2cm} {\phantom{\huge{j}}}
     &&\qquad$\mu_{\rm sea 2}$ & 1.807(44) &  2.029(49) &  0.268(7)   & 0.336(15)\\ 
 {\phantom{\huge{l}}} \raisebox{-.2cm} {\phantom{\huge{j}}}
     &&\qquad$\mu_{\rm sea 3}$      & 1.798(44) &  2.020(49) &  0.263(6)  & 0.330(10)\\ 
{\phantom{\huge{l}}} \raisebox{-.2cm} {\phantom{\huge{j}}}
     &&\qquad$\mu_{\rm sea 4}$    & 1.810(44)  & 2.029(49) & 0.272(7)  & 0.333(15)\\ \hline
{\phantom{\huge{l}}} \raisebox{-.2cm} {\phantom{\huge{j}}}
3.90 &  \qquad 32 &\qquad$\mu_{\rm sea 1}$ &  1.803(44) & 2.022(48) &  0.266(6)  & 0.331(10)\\ 
{\phantom{\huge{l}}} \raisebox{-.2cm} {\phantom{\huge{j}}}
    & &\qquad $\mu_{\rm sea 2}$ &  1.803(44)& 2.019(49) & 0.267(6)  & 0.330(10)\\ \hline
{\phantom{\huge{l}}} \raisebox{-.2cm} {\phantom{\huge{j}}}
4.05 & \qquad 32&\qquad $\mu_{\rm sea 1}$ & 1.877(27) & 2.099(31) & 0.264(5) &  0.341(9)\\ 
{\phantom{\huge{l}}} \raisebox{-.2cm} {\phantom{\huge{j}}}
     &&\qquad $\mu_{\rm sea 2}$  & 1.878(27)& 2.079(32) & 0.267(5) &  0.331(9)\\ 
{\phantom{\huge{l}}} \raisebox{-.2cm} {\phantom{\huge{j}}}
     &&\qquad $\mu_{\rm sea 3}$ &  1.880(27) & 2.102(30) &  0.267(6)  & 0.339(9)\\ \hline
{\phantom{\huge{l}}} \raisebox{-.2cm} {\phantom{\huge{j}}}
4.20 & \qquad 32& \qquad $\mu_{\rm sea 1}$ &  1.909(38) & 2.104(39) &  0.266(5) &  0.320(10)\\  \hline
{\phantom{\huge{l}}} \raisebox{-.2cm} {\phantom{\huge{j}}}
4.20      & \qquad 48 &\qquad $\mu_{\rm sea 1}$   & 1.888(39) &  2.096(38)  & 0.251(5) &  0.316(8)
\\ \hline
\end{tabular}}}
{\caption{\footnotesize  \label{tab:2} 
Charmed meson decay constants and hadron masses with the light quark fixed to the strange quark mass~\cite{Blossier:2010cr}. All results are given in GeV.}}
\end{table}

\begin{figure}
\begin{center}
\begin{tabular}{@{\hspace{-0.25cm}}c}
\epsfxsize12.7cm\epsffile{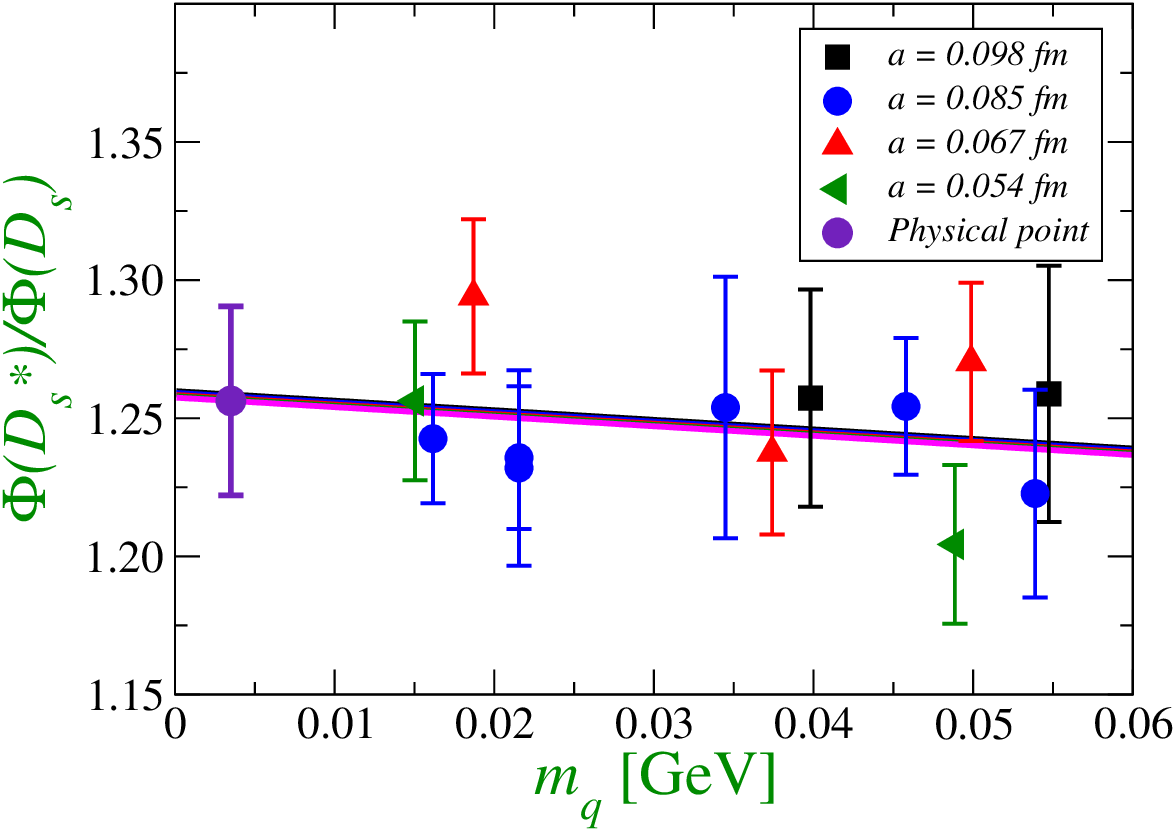}    \\
\end{tabular}
\vspace*{-.1cm}
%%%%%%%%%%%%%%%%%%%%%%%%%%%%%%%%%%%%%%%%%%%%%%%%%%%%%%%%%%%%%%%%%%
\caption{\label{fig:3}{\footnotesize 
Dependence of the ratio of  $\Phi({D_s^\ast})/\Phi(D_s)$ on the sea quark mass, where $m_q \equiv m_q^\msbar (2\ \gev)$. The lattice discretization errors appear to be negligible. 
 } }
%%%%%%%%%%%%%%%%%%%%%%%%%%%%%%%%%%%%%%%%%%%%%%%%%%%%%%%%%%%%%%%%%%
\end{center}
\end{figure}

In tab.~\ref{tab:2} we give the values for the masses and decay constants as obtained in each of the ensembles of gauge field configurations described in tab.~\ref{tab:1}. We should stress that the charm and strange quark mass were obtained from completely separated studies~\cite{Blossier:2010cr}, and that the values of vector meson masses and decay constants are the net prediction of lattice QCD. For the reader's convenience we converted our results from lattice units to the physical ones by using the lattice spacing values given in tab.~\ref{tab:1}. From tab.~\ref{tab:2} we also see that the errors on the  vector meson decay constants are twice larger than for the pseudoscalar ones. Finally, to get to the physical result we need to extrapolate in the sea quark from the masses used in the simulations down to the physical $u/d$-quark mass in the continuum limit. To do so we combine all results obtained with the valence physical charm and strange quarks,  and fit our data to
\bea\label{r3}
\Phi({D_s^\ast})^{\rm latt.} =\left(f_{D_s^\ast} \sqrt{m_{D_s^\ast}}\right)^{\rm phys.} \left[ 1\  +\  A_s m_q^{\rm sea} \  + \ B_s a^2 \right] \,,
\eea
where on the left hand side we combine the masses and decay constants into the quantity $\Phi(D_s^\ast) =f_{D_s^\ast} \sqrt{m_{D_s^\ast}}$. $A_s$ and $B_s$ are the fit parameters. One can use the same formula to extrapolate the decay constant alone. An expression similar to eq.~(\ref{r3}) is also used for the pseudoscalar decay constant, and so one can estimate the size of the heavy quark spin symmetry breaking effects in the decay constants through the ratio $\Phi(D_s^\ast)/\Phi(D_s)$.   Clearly, from our data, shown in fig.~\ref{fig:3}, we do not see any sea quark mass dependence.  Furthermore the ${\cal O}(a^2)$ effects are very small, and we finally obtain
%%%%%%%%%%%%%%%%%%%%%%%%%%%%%%%%%%%%%%%%%%%%%%%%%%%%%%%%%%%
\bea\label{r-6}
f_{D^\ast_s} = 311\pm 9~\mev\,, \quad {f_{D^\ast_s} \over f_{D_s}}  = 1.26\pm 0.03\,.
\eea
Our result in eq.~(\ref{r-6}) shows that the heavy quark spin symmetry breaking effects are larger than $20\%$ for the case in which the heavy quark is charm. This simply means that for the decay constants the charm physics is far away from the static limit in which $\displaystyle{\lim_{m_c\to \infty}} (f_{D^\ast_s}/f_{D_s}) = 1$, and that power corrections, proportional to $1/m_c^n$, are important.  To better appreciate this fact we remind the reader that $f_{\rho^+}/f_{\pi^+}\simeq 1.58$, and $f_{K^{\ast +}}/f_{K^+}\simeq 1.36$ (see for example \cite{PDG,FT}).  Note also that both $f_{D^\ast_s}$ and ${f_{D^\ast_s}/f_{D_s}}$ are somewhat larger than the results obtained in the quenched approximation~\cite{quenched-D}. 
We should also add that from a similar chiral and continuum extrapolation of the meson mass we obtain $m_{D^\ast_s} = 2141(22)$~MeV, compatible with the experimentally established $m^{\rm PDG}_{D^\ast_s} = 2112$~MeV~\cite{PDG}.

\subsection{$f_{D^\ast}$,  $f_{D^\ast}/f_D$, ${f_{D^\ast_s}/f_{D^\ast}}$, and other ratios}
Next we present our results for masses and decay constants of charmed mesons containing a light (up or down) quark. We treat the light quark mass as fully unquenched, i.e. we keep the light valence and sea quarks degenerate in mass. 
\begin{table}[h!!]
\centering 
\begin{tabular}{|l cc|c|c|c|c|cc|cc|}  \hline
{\phantom{\huge{l}}} \raisebox{-.2cm} {\phantom{\huge{j}}}
$\beta$ & $L$  & \qquad $\mu_{\rm sea }$   & $ m_{D_q}$            & $ m_{D^\ast_q} $       & $f_{D_q}$          & $f_{D^\ast_q}$       \\ \hline\hline
{\phantom{\huge{l}}} \raisebox{-.2cm} {\phantom{\huge{j}}}
3.80 & 24 & \qquad$\mu_{\rm sea 1}$ & 1.722(52) &1.946(58)  & 0.264(9)  & 0.321(19)\\  
{\phantom{\huge{l}}} \raisebox{-.2cm} {\phantom{\huge{j}}}
    & &\qquad$\mu_{\rm sea 2}$ & 1.735(52) & 1.967(57) & 0.267(9) & 0.338(19) \\ \hline
{\phantom{\huge{l}}} \raisebox{-.2cm} {\phantom{\huge{j}}}
3.90 & 24 & \qquad$\mu_{\rm sea 1}$ & 1.741(44)  & 1.969(48)  & 0.244(8)  & 0.311(21) \\ 
{\phantom{\huge{l}}} \raisebox{-.2cm} {\phantom{\huge{j}}}
  &   &\qquad$\mu_{\rm sea 2}$ & 1.751(43)  & 1.970(51)  & 0.251(7)  & 0.313(21) \\  
{\phantom{\huge{l}}} \raisebox{-.2cm} {\phantom{\huge{j}}}
   &  &\qquad$\mu_{\rm sea 3}$ & 1.748(43) &  1.967(49) &  0.248(6) &  0.311(11) \\
{\phantom{\huge{l}}} \raisebox{-.2cm} {\phantom{\huge{j}}}
   &  &\qquad$\mu_{\rm sea 4}$ & 1.774(42) &  1.992(51) &  0.262(7) &  0.317(21) \\ \hline
{\phantom{\huge{l}}} \raisebox{-.2cm} {\phantom{\huge{j}}}
3.90  & 32 & \qquad$\mu_{\rm sea 1}$ & 1.733(43)  & 1.958(51) & 0.244(7) & 0.312(20) \\ 
{\phantom{\huge{l}}} \raisebox{-.2cm} {\phantom{\huge{j}}}
    &  &\qquad $\mu_{\rm sea 2}$ & 1.732(42)  & 1.935(48)  & 0.244(7) &  0.298(11) \\ \hline
{\phantom{\huge{l}}} \raisebox{-.2cm} {\phantom{\huge{j}}}
4.05  & 32 &\qquad $\mu_{\rm sea 1}$ & 1.808(27)& 2.024(31) &  0.241(7) &  0.312(12) \\ 
{\phantom{\huge{l}}} \raisebox{-.2cm} {\phantom{\huge{j}}}
  &   &\qquad $\mu_{\rm sea 2}$ & 1.823(27)  & 2.026(33)  & 0.250(6)  & 0.313(11) \\ 
{\phantom{\huge{l}}} \raisebox{-.2cm} {\phantom{\huge{j}}}
    & &\qquad $\mu_{\rm sea 3}$ & 1.840(27)  & 2.067(30) & 0.256(8)  & 0.328(10) \\ \hline
{\phantom{\huge{l}}} \raisebox{-.2cm} {\phantom{\huge{j}}}
4.20  & 32 &\qquad $\mu_{\rm sea 1}$ & 1.866(35) & 2.064(38) & 0.252(6) & 0.306(12) \\ \hline
{\phantom{\huge{l}}} \raisebox{-.2cm} {\phantom{\huge{j}}}
 4.20  & 48    &\qquad $\mu_{\rm sea 1}$ & 1.806(37)  & 2.032(42)  & 0.224(7)  & 0.290(15) 
\\ \hline
\end{tabular}
{\caption{\footnotesize  \label{tab:3} 
Charmed non-strange meson masses and decay constants for various light quark masses. All results are obtained with the light valence quarks mass degenerate with the sea quark mass. Like in tab.~\ref{tab:2}, all numbers are given in physical units [GeV]. }}
\end{table}
As before, and for the readers' convenience, we list all our results in physical units and together with the corresponding  masses and decay constants of the pseudoscalar mesons. As in the previous subsection, we proceed by combining all our data to make the combined chiral and continuum extrapolation via the expression, 
\bea\label{chi-fit}
\Phi(D_q^\ast)^{\rm latt.} = \left( f_{D^\ast} \sqrt{ m_{D^\ast} } \right)^{\rm phys} \left[ 1 - {3 \over 4} { 1+3 g^2 \over  (4\pi f_\pi)^2} m_\pi^2 \log m_\pi^2 + A_{u,d} m_\pi^2+   B_{u,d} a^2\right]\,,
\eea
where again we considered the quantity $f_{D^\ast_q} \sqrt{ m_{D^\ast_q}}=\Phi(D_q^\ast)$ for which the above formula, derived in heavy meson chiral perturbation theory (HMChPT), can be used to guide the chiral extrapolation~\cite{HMChPT}. Since this step turns out to be the dominant source of systematic uncertainties, we should spend a few more lines to discuss it. Despite the fact that the window of light quark masses used in our data stretches down to about $1/7$ of the strange quark mass, the fact that the chiral logarithmic correction to this quantity is large introduces an important uncertainty in the result of the extrapolation. First of all, when applying the above expression to the data obtained with the charmed heavy quark, it is not clear what value for the coupling of the heavy-light mesons to a soft pion, $g$, to use. While its value in the static heavy quark limit ($m_c\to \infty$) has been computed on the lattice and was shown to be $g\simeq 0.4\div 0.5$~\cite{g-hat}, its value computed with the propagating charm quark was found to be $g\simeq 0.7$~\cite{Becirevic:2009xp}. Since the chiral corrections were computed in the static heavy quark limit one is tempted to use the value for $g$ obtained in the static limit as well. However, knowing that the $1/m_c^n$-corrections are sizable, the validity of HMChPT in the charm sector is dubious and one should try other possibilities, including the larger $g\simeq 0.7$, or even the simple linear chiral extrapolation. Another aspect, that should not be ignored, is the fact that the HMChPT formulas were obtained by taking into account the lowest doublet of heavy-light mesons only. As we know, however, the first orbital excitations (scalar and axial mesons) turned out to be very close to the lowest lying states (pseudoscalar and vector mesons), and their inclusion might modify the chiral extrapolations as well~\cite{use-misuse}. To make a fair assessment of the chiral uncertainties, each of our central values will correspond to the result obtained by using the expression~(\ref{chi-fit}) with $g=0.45$, and the difference between that and: ({\it a}) the result obtained by using $g=0.7$ in eq.~(\ref{chi-fit}), ({\it b}) the result obtained through the linear  chiral extrapolation, will be added separately as an uncertainty to our result.  To give the reader a good feeling about the smoothness of the chiral and continuum extrapolations, here are the numbers obtained from the fit to our data,
\bea\label{a-1}
 \Phi(D_q^\ast):\;  f_{D^\ast} \sqrt{ m_{D^\ast} } = 382(2)~\mev^{3/2}\,,   A_{u,d} =0.0(3)~\gev^{-2}\,,   B_{u,d} = 0.0(3)~{\rm fm}^{-2}\,,
\eea
which resemble very much the shape of the combined chiral and continuum extrapolations in the case of pseudoscalar mesons, namely,
\bea\label{a-2}
 \Phi(D_q):\;  f_{D} \sqrt{ m_{D} } = 276(8)~\mev^{3/2}\,,   A^\prime_{u,d} = 0.0(2)~\gev^{-2}\,,   B^\prime_{u,d} = 0.0(2)~{\rm fm}^{-2}\,.
\eea
 It goes without saying that the chiral logarithmic correction derived in HMChPT in the static limit is the same for both the pseudoscalar and vector mesons. 

What might look striking from the above numbers is the fact that the discretization errors are essentially absent. 
This is actually a fortunate artifact of the quantities  $\Phi(D_q^\ast)$ and  $\Phi(D_q)$. 
The results listed in tab.~\ref{tab:3} obviously show the non-negligible $a^2$-effects. The decay constant exhibits a positive slope in $a^2$ and the continuum value for $f_{D^{(\ast)}}$ is smaller 
than those obtained at fixed value of the lattice spacing. More pronounced slope in $a^2$  shows up in the results  for the heavy-light meson masses, but of the sign opposite to that in the decay constants. Those two effects cancel against each other to a large extent, which is why $\Phi(D^{(\ast)})=f_{D^{(\ast)}} \sqrt{ m_{D^{(\ast)}}}$ is manifestly insensitive to discretization effects  (the parameter $B^{(\prime)}_{u,d}$ obtained from our fit is consistent with zero.) 

Our final results are:
\bea
f_{D^\ast} = 275\pm 13 {}^{+12}_{-07}~\mev\,, \quad {f_{D^\ast} \over f_{D}}  = 1.28\pm 0.06\,.
\eea
where the second error corresponds to the uncertainty arising from the chiral extrapolation: the upper error is the difference with the central value of the linear extrapolation, and the lower is the difference with result obtained by using $g=0.7$ instead of $g=0.45$ in eq.~(\ref{chi-fit}).  As for the $D^\ast$ meson mass, we obtain $m_{D^\ast}=2041(22)$~MeV, that is again somewhat larger than the measured $m_{D^\ast}^{\rm PDG}=2010$~MeV. We should mention that the previous studies with tmQCD on the lattice observed that the light vector meson masses are larger than the physical ones~\cite{Dimopoulos:2011cf,Jansen:2009hr}.  Although to a much lesser extent, this feature seems to persist  with the heavy-light mesons too.~\footnote{For example in ref.~\cite{Dimopoulos:2011cf} the authors quote $m_{K^\ast} = 0.981(33)$~GeV, the central value of which is about $10$\% larger than $m_{K^\ast}^{\rm PDG}=0.892$~GeV. In our case,  the central value for our  $m_{D^\ast}$ is $2$\% larger then the experimentally measured mass. This result is obtained by using the smeared source operators. The non-smeared (local) sources give a value compatible with the smeared one, within the error bars. Its central value is however a tad larger and is $3.5$\% higher than the experimentally established $D^\ast$-meson mass. }
\begin{figure}
\vspace*{-0.8cm}
\begin{center}
\begin{tabular}{@{\hspace{-0.25cm}}c}
\epsfxsize12.7cm\epsffile{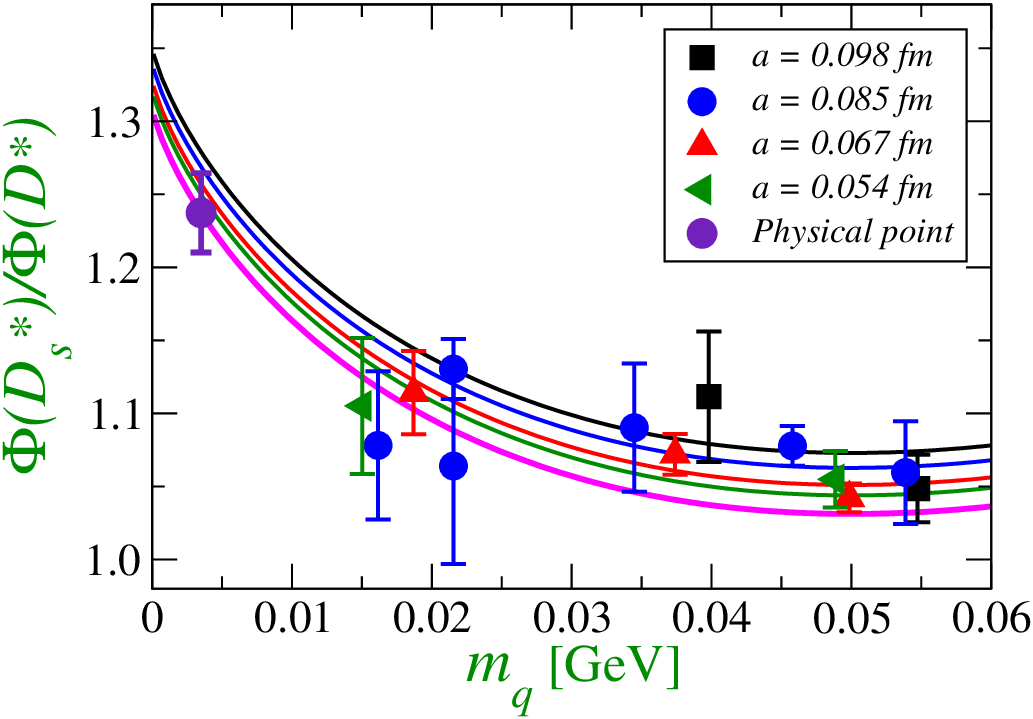}    \\
\end{tabular}
\vspace*{-.1cm}
%%%%%%%%%%%%%%%%%%%%%%%%%%%%%%%%%%%%%%%%%%%%%%%%%%%%%%%%%%%%%%%%%%
\caption{\label{fig:4}{\footnotesize 
SU(3) breaking ratio for the charmed vector meson decay constants. Dashed curves correspond to the chiral extrapolation 
using eq.~(\ref{chi-fit}) for each lattice spacing and their spread indicates the size of discretization effects. The full thick curve is the result of chiral extrapolation in the continuum limit. 
 } }
%%%%%%%%%%%%%%%%%%%%%%%%%%%%%%%%%%%%%%%%%%%%%%%%%%%%%%%%%%%%%%%%%%
\end{center}
\end{figure}
%%%%%%%%%%%%%%%%%%%%%%%%%%%%%%%%%%%%%%%%%%%%%%%%%%%%%%%%%%%

Other interesting and phenomenologically useful ratios are:
\bea\label{r-9}
 &&{f_{D^\ast_s} \over f_{D^\ast}}  = 1.17\pm 0.02 {}^{+05}_{-07}\,,\quad  {f_{D_s} \over f_{D}}  = 1.19\pm 0.01 {}^{+05}_{-07}\,,\nn\\
 && {f_{D_s} \over f_{D^\ast}}  = 0.90\pm 0.02 {}^{+02}_{-04}\,, \quad   {f_{D^\ast_s} \over f_{D}}  = 1.48\pm 0.05 {}^{+05}_{-08}\,.
\eea
Note, in particular,  that our results are consistent with the recently claimed inequality $ {f_{D^\ast_s}/f_{D^\ast}} <  {f_{D_s}/f_{D}}$ deduced from HMChPT, after including the power corrections and assuming the validity of $SU(3)$ chiral perturbation theory for the light quarks~\cite{Altenbuchinger:2011qn}. 

\section{\label{sec-4}Verifying the factorization approximation}

With the information on the charmed meson decay constants in hands, we can now illustrate on the specific examples how one can check the validity of the factorization approximation in several particular non-leptonic $B$-decays mentioned in introduction of the present paper. To begin with, we remind the reader that the following modes have been measured~\cite{PDG}:
\begin{align}\label{pdg-values}
B(B^0\to D^+\pi^-) &= \left( 4.6 \pm 0.4 \right) \times 10^{-5}, &B(B^0\to D^+D^-) = \left( 2.11 \pm 0.31 \right) \times 10^{-4}\,,\nn\\
B(B^0\to D^{\ast +}\pi^-)& = {\rm N.A.}, &B(B^0\to D^{\ast +}D^-) = \left( 6.1 \pm 1.5 \right) \times 10^{-4}\,,\nn\\
B(B^0\to D_s^+\pi^-) &= \left( 2.16 \pm 0.26 \right) \times 10^{-5}, &B(B^0\to D_s^+D^-) = \left( 7.2 \pm 0.8 \right) \times 10^{-3}\,,\nn\\
B(B^0\to D_s^{\ast +}\pi^-)& = \left( 2.1 \pm 0.4 \right) \times 10^{-5}, &B(B^0\to D_s^{\ast +}D^-) = \left( 7.4 \pm 1.6 \right) \times 10^{-3}\,.\nn\\
\end{align}
We consider the following four ratios:
\begin{align}
R_1&=&{B(B^0\to D_s^+\pi^-)\over B(B^0\to D^+\pi^-)} = \left({V_{cs}\over V_{cd}}\right)^2  \, \left[ {\lambda(m_B,m_{D_s},m_{\pi})\over \lambda(m_B,m_{D},m_{\pi})}\right]^{1/2}  \, 
\left[  {F^{B\to \pi}_0 (m_{D_s}^2) \over F^{B\to \pi}_0 (m_{D}^2)} \right]^2
\left(  {f_{D_s} \over f_{D}} \right)^2\,,\nn\\
R_2&=&{B(B^0\to D_s^+D^-)\over B(B^0\to D^+D^-)} = \left({V_{cs}\over V_{cd}}\right)^2  \, \left[ {\lambda(m_B,m_{D_s},m_{D})\over \lambda(m_B,m_{D},m_{D})}\right]^{1/2}  \left[  {F^{B\to D}_0 (m_{D_s}^2) \over F^{B\to D}_0 (m_{D}^2)} \right]^2 
\left(  {f_{D_s} \over f_{D}} \right)^2\,,\nn
\end{align}
\begin{align}
\label{eq-X}
R_3&=&{B(B^0\to D_s^{\ast +}\pi^-)\over B(B^0\to D^{\ast +}\pi^-)} = \left({V_{cs}\over V_{cd}}\right)^2  \, \left[ {\lambda(m_B,m_{D_s^\ast},m_{\pi})\over \lambda(m_B,m_{D^\ast},m_{\pi})}\right]^{3/2} 
\left[  {F^{B\to \pi}_+ (m_{D_s^\ast}^2) \over F^{B\to \pi}_+ (m_{D^\ast}^2)} \right]^2 
 \left(  {f_{D_s^\ast} \over f_{D^\ast}} \right)^2\,,\nn\\
R_4&=&{B(B^0\to D_s^{\ast +}D^-)\over B(B^0\to D^{\ast +}D^-)} = \left({V_{cs}\over V_{cd}}\right)^2  \, \left[ {\lambda(m_B,m_{D_s^\ast},m_{D})\over \lambda(m_B,m_{D^\ast},m_{D})}\right]^{3/2}
\left[  {F^{B\to D}_+ (m_{D_s^\ast}^2) \over F^{B\to D}_+ (m_{D^\ast}^2)} \right]^2 
 \left(  {f_{D_s^\ast} \over f_{D^\ast}} \right)^2\,,
\end{align}
where on the right hand side we explicitly write the expressions obtained in the factorization approximation, with the usual $\lambda(a,b,c)=[a^2-(b-c)^2][a^2-(b+c)^2]$.  Ratios of the semileptonic form factors on the right hand side can be taken to be one to excellent accuracy, so that by using our results from eq.~(\ref{r-6}), together with $V_{cs}/V_{cd}=1/\tan\theta_C =4.32$, we obtain:
\begin{align}\label{r1234}
&R_1^{\rm (fact.)} = 26.0\pm 0.4\pm 2.6 \,,\quad {\rm vs.} \quad R_1^{\rm (exp.)} = 0.47\pm 0.07\,, &\nn \\
&R_2^{\rm (fact.)} = 25.7\pm 0.4\pm 2.6 \,,\quad {\rm vs.} \quad R_2^{\rm (exp.)} = 34.1\pm 6.3\,, &\nn  \\
&R_3^{\rm (fact.)} = 23.8\pm 0.8\pm 2.5 \,,\quad {\rm vs.} \quad R_3^{\rm (exp.)} = {\rm N.A.}\,, &\nn  \\
&R_4^{\rm (fact.)} = 22.7\pm 0.8\pm 2.4 \,,\quad {\rm vs.} \quad R_4^{\rm (exp.)} = 12.1\pm 4.0\,. &
\end{align}
The first ratio shows a very large disagreement between the factorization approximation and experiment, and it is mainly due to $B^0\to D^+\pi^-$ for which the experimental branching fraction is by almost two orders of magnitude larger than the one obtained in the factorization approximation. To understand the origin of this discrepancy we looked in the original paper by the Belle Collaboration~\cite{das}, and realized that the value for $B(B^0\to D^+\pi^-)$ quoted by PDG, and given in eq.~(\ref{pdg-values}) above, is not actually measured but deduced from the measurement of $B(B^0\to D_s^+\pi^-)$, and by imposing the validity of factorization, similar to our ratio $R_1$ above. Furthermore, PDG erroneously identified the measured value of $R_{D\pi}=1.71(11)(9)(2)\%$~\cite{das}, with $R_{D\pi} = B(B^0\to D^+\pi^-)/B(B^0\to D^-\pi^+)$, while in fact the definition given in ref.~\cite{das}, after assuming the factorization, reads $R_{D\pi} = \sqrt{B(B^0\to D^+\pi^-)/B(B^0\to D^-\pi^+)}$.  We corrected that error, and by using $B(B^0\to D^-\pi^+)= (2.68\pm 0.13) \times 10^{-3}$~\cite{PDG}, we obtain 
\bea\label{pdg-x}
B(B^0\to D^+\pi^-)^{\rm PDG}= (7.8 \pm 1.4) \times 10^{-7}. 
\eea
Since the value of  $B(B^0\to D^+\pi^-)$ is not measured independently but extracted by imposing the factorization approximation, the comparison of our $R_1^{\rm (fact.)}$ with 
corrected $R_1^{\rm (exp.)}$ would not provide us with any useful information. Instead, we can follow the same recipe as PDG, and use our value for $R_1^{\rm (fact.)}$ from eq.~(\ref{r1234}), to arrive at  
\bea\label{ourPImoins}
B(B^0\to D^+\pi^-) =  \left( 8.3 \pm 1.0 \pm 0.8 \right) \times 10^{-7}\,,
\eea
where the first error is the experimental uncertainty coming from $B(B^0\to D_s^+\pi^-)$, and the second is our theoretical uncertainty within the factorization approximation. The reason why the results in eqs.~(\ref{pdg-x}) and (\ref{ourPImoins}) differ is the fact that in ref.~\cite{das} the value of SU(3) breaking ratio $f_{D_s}/f_D=1.164(11)$ has been used, while our value is  $f_{D_s}/f_D=1.18(1)(6)$. Furthermore, in converting our $R_1^{\rm (fact.)}$ to the result~(\ref{ourPImoins}) we used $B(B^0\to D_s^+\pi^-)$ from ref.~\cite{PDG}, which is the average of the results obtained by Belle~\cite{das} and BaBar~\cite{aubert} .

Similarly, our result for the ratio $R_3^{\rm (fact.)}$ can be combined with the measured $B(B^0\to D_s^{\ast +}\pi^-)$ to deduce, 
\bea
B(B^0\to D^{\ast +}\pi^-) =(8.8\pm 1.6 \pm  0.9)\times 10^{-7}\,,
\eea 
where the first error reflects the experimental uncertainty in $B(B^0\to D_s^{\ast +}\pi^-)$, and the second is theoretical error within the factorization approximation.

In the case with two charmed mesons in the final state, we see that the factorization approximation works rather well. The difference between $R_{2,4}^{\rm (fact.)} $ and $R_{2,4}^{\rm (exp.)}$ can be used to constrain the corrections to the factorization coming from either the penguin (Cabibbo suppressed) contributions or the final state interaction~\cite{xxxxx}. 

To use other results from eqs.~(\ref{r-6},\ref{r-9}) and check whether or not the factorization approximation in ratios of various decay modes agrees with experiment, one also needs different form factors, and an extra assumption is needed. For example, 
\bea
R_5&=&{B(B^0\to D_s^{\ast +}\pi^-)\over B(B^0\to D_s^+\pi^-)} \stackrel{\rm fact.}{=} {  \left[ \lambda(m_B,m_{D_s^\ast},m_{\pi})\right]^{3/2} \over (m_B^2-m_\pi^2)^2  \left[\lambda(m_B,m_{D_s},m_{\pi})\right]^{1/2}} \  \left[  {F^{B\to \pi}_+ (m_{D_s^\ast}^2) \over F^{B\to \pi}_0 (m_{D_s}^2)} \right]^2 
\left(  {f_{D_s^\ast} \over f_{D_s}} \right)^2\,,\nn\\
&=& \left( 1.09 \pm 0.05\right) \   \left[  {F^{B\to \pi}_+ (m_{D_s^\ast}^2) \over F^{B\to \pi}_0 (m_{D_s}^2)} \right]^2\,,
\eea
where, in the last line, we used our result from eq.~(\ref{r-6}). Knowing that the two semileptonic $B\to \pi$ form factors are slowly varying functions at low $q^2$'s, and that $F_+ (0)=F_0 (0)$, it is reasonable to assume that 
${F^{B\to \pi}_+ (m_{D_s^\ast}^2) \approx F^{B\to \pi}_0 (m_{D_s}^2)}$, and therefore:
\bea
R_5^{\rm (fact.)} \simeq 1.09\pm 0.05 \,,\quad {\rm vs.} \quad R_5^{\rm (exp.)} = 1.0\pm 0.2\,,
\eea 
a comparison that again goes in favor of the validity of factorization. 

\section{\label{sec-5}Summary}
In this paper we discussed the extraction of the charmed vector meson decay constants from the simulations of tmQCD on the lattice with $N_{\rm f}=2$ dynamical light quarks. The results we obtain are presented along with those of the pseudoscalar mesons, clearly showing that the heavy quark spin symmetry breaking effects are large in the case of the heavy charm quark, i.e. that corrections $\propto 1/m_c^n$ are sizable. The dominant sources of errors are those coming from the chiral extrapolations when the extrapolation in the valence light quark mass was needed (i.e. for $D$-mesons). We do not make a `guestimate' of the systematic uncertainty that might arise from the non-included strange and charm quark in the sea. That point will be numerically assessed once the ongoing simulations with $N_{\rm f}=2+1+1$ dynamical quark flavors are completed. As for the main  numerical results presented in this paper, after symmetrizing the error bars, we quote:
\bea
&&f_{D^\ast} = 278\pm 13 \pm 10~\mev\,, \quad f_{D^\ast_s} = 311\pm 9~\mev\,\nn\\
&&\hfill \nn\\
&& {f_{D^\ast_s} \over f_{D_s}}  = 1.26\pm 0.03\,, \quad{f_{D^\ast} \over f_{D}}  = 1.28\pm 0.06\,, \quad\, {f_{D^\ast_s} \over f_{D^\ast}}  = 1.16\pm 0.02\pm 0.06\,.
\eea
In some cases the results obtained in this paper can be used to verify the validity of the factorization approximation in non-leptonic $B$-decay modes. We also corrected the error in PDG and instead of the currently reported $B(B^0\to D^+\pi^-) = \left( 4.6 \pm 0.4 \right) \times 10^{-5}$, we obtain $B(B^0\to D^+\pi^-) =  \left( 7.8 \pm 1.4 \right) \times 10^{-7}$, which is in very good agreement with the value we obtain from another ratio [$R_1$ in eq.~(\ref{eq-X})], namely $B(B^0\to D^+\pi^-) =  \left( 8.3 \pm 1.0 \pm 0.8 \right) \times 10^{-7}$. Both values are obtained by relying on the validity of the factorization approximation. 
 
\vspace{1.7 cm}

\section*{Acknowledgments}
We thank M.~Ciuchini, A.~LeYaouanc and L.~Oliver for comments, all the members of the ETM Collaboration for many discussions and for making their gauge field configurations publicly available. V.L., S.S. and C.T. thank MIUR (Italy) for partial financial support under the
contracts PRIN08. Partial support by the EU contract HP2-227431 is kindly acknowledged as well.

\end{document}